\documentclass[12pt,graphicx,subfigure,axodraw]{article}
\usepackage{amsfonts}
\usepackage{amssymb}
\usepackage[usenames,dvipsnames]{color}
\usepackage{graphicx}

\newcommand{\beqn}{\begin{eqnarray}}
\newcommand{\eeqn}{\end{eqnarray}}
\newcommand{\be}{\begin{equation}}
\newcommand{\ee}{\end{equation}}
\newcommand{\beq}{\begin{equation}}
\newcommand{\eeq}{\end{equation}}

\def\s1{$s_{\alpha}$}
\def\s2{$s_{\gamma}$}
\def\s3{$s_{\delta}$}
\def\c1{$c_{\alpha}$}
\def\c2{$c_{\gamma}$}
\def\c3{$c_{\delta}$}

\def\br{\left(\begin{array}{c}}
\def\er{\end{array}\right)}
\begin{document}
\baselineskip 18pt

\thispagestyle{empty}

\vspace{0cm}

\begin{center}
{\bf  \Large
{The Chromoelectric Dipole Moment  of the Top Quark 
in Models with  Vector Like  Multiplets } }
\vspace{1.0cm}

{\bf Tarek Ibrahim}\footnote{e-mail: tarek-ibrahim@alex-sci.edu.eg}$^{,a}$ and 
{\bf Pran Nath}\footnote{e-mail: nath@neu.edu}$^{,b}$
\\
\vspace{.5cm}

{\it
$^{a}$  Department of  Physics, Faculty of Science,
University of Alexandria, Alexandria, Egypt\\ 
$^{b}$Department of Physics, Northeastern University,
Boston, Massachusetts 02115, USA \\

}

\end{center}

\vspace{0.2cm}

\begin{center}
{\bf Abstract} \\
\end{center}
\vspace{0cm}
The  chromoelectric dipole moment of the  top quark is calculated in a model with a vector
like multiplet which mixes with the third generation in an extension of the MSSM. Such mixings allow for new CP violating phases.
Including these new CP phases, the chromoelectric dipole moment that generates an electric dipole
 of the  top  in this class of models is computed. The top chromoelectric dipole moment operator
arises from loops
involving the exchange of the W, the Z  
 as well as from the exchange involving  the charginos, the neutralinos, the gluino, and the  
 vector like multiplet and their superpartners.
 The analysis of the chromoelectric dipole moment operator
 of the top is more complicated than for the light quarks 
 because the mass of the external fermion,
 in this case the  top quark mass,  cannot be ignored relative to the masses inside the loops. 
A numerical analysis is presented and it is shown that the contribution to the top EDM could 
lie in the range 
 ($10^{-19}-10^{-18})$ ecm  
consistent with the current limits on the EDM of the electron, the neutron and on atomic EDMs.
A top EDM of size $(10^{-19}-10^{-18})$ ecm could be accessible in collider experiments such as 
at the LHC and at  the ILC.
\setcounter{footnote}{0}
\clearpage

\section{Introduction\label{1}}

  The electric dipole moment  (EDM) of elementary particles provide an important window to possible 
  new sources of  CP violation (For recent reviews see\cite{Ibrahim:2007fb}).  This is so because
  in the Standard Model the EDM of an elementary particle is rather small. Thus for the top quark
  the EDM in the Standard Model is estimated to be less than $10^{-30}$ ecm\cite{Hoogeveen:1990cb,Soni:1992tn,note}   and  outside the realm of experiment in the foreseeable future (For a review of CP violation in top physics 
  see \cite{Atwood:2000tu}).
  However, much larger EDMs for elementary particles can arise in new physics models.
  One such model considered recently was  where one has extra vector like generations
  which can mix with the third generation\cite{Ibrahim:2008gg,Ibrahim:2010va,Ibrahim:2010hv}.
  Extra vector like generations can arise in many unified theories of particle physics\cite{Georgi:1979md,Senjanovic:1984rw}  
  and if their masses lie in the TeV range they could mix with the third  generation and produce
  observable effects. Such mixings are consistent with the current precision electroweak data\cite{Jezabek:1994zv}
  and thus the implications of such vector like multiplets have been analyzed in a number of
   works\cite{Barger:2006fm,Lavoura:1992qd,Maekawa:1995ha,Morrissey:2003sc,Choudhury:2001hs,Liu:2009cc,Babu:2008ge,Martin:2009bg,Graham:2009gy}.  In \cite{Ibrahim:2010hv} an analysis of the electric dipole operator
   for the top quark was given    arising from the exchange of the extra vector like generations in the loops
   and it was found that a significantly larger EDM 
    than in the Standard Model can arise for the top quark
    from  such exchanges. In this work we analyze the contribution to the chromoelectric dipole operator (CEDM)
    from the exchange of the vector like generations in the loops. Our analysis is done in
    an extension of the minimal supersymmetric standard model (MSSM) including the extra vector like multiplets. 
     The analysis shows that  a top  EDM as large 
 as $(10^{-19}-10^{-18})$ ecm can
     arise from a constructive interference between the electric dipole moment and the chromoelectric dipole 
     moment. A top EDM of this size lies   within the realm of future experiment \cite{Frey:1997sg,Atwood:1992vj,Poulose:1997xk,Choi:1995kp}.  
  The role of EDMs in a variety of processes 
  such as $e^+e^-\to t\bar t$, $\gamma\gamma \to t\bar t$ and other phenomena have been investigated
  by a number of  authors 
  \cite{Atwood:1992vj,as2,Bartl:1997wf,Hollik:1998vz,Hollik:1998vz,NovalesSanchez:2009zz,Huang:1994zg}
  and thus the EDM of the top is of significant interest.

   The outline of the  rest of the paper is as follows:
   In Sec.(\ref{2}) we define the chromoelectric dipole 
   moment of the quark and its connection with the electric dipole moment.
       In Sec.(\ref{3}) we  give an analysis of the
 EDM of the top 
 allowing for mixing between the vector like multiplet  and the
third generation quarks in the underlying model discussed in \cite{Ibrahim:2010hv}. These mixings contain new sources of CP violation. 
 Here we compute the loops
involving the exchanges of the W and the
 Z,  of the  charginos, of the neutralinos, of the gluino as
well as exchanges involving the vector like multiplets and their superpartners. 
In Sec.(\ref{4}) we discuss the parameter space of the model and list the new CP violating 
phases that enter in the analysis. 
A numerical analysis of the size of the EDM of the top is given in Sec.(\ref{5}). 
In this section we also  display the  dependence of the top EMD on the  CP phases  
arising from the mixings of the third generation quarks  with the extra vector like generations.  
Conclusions are given in Sec.(\ref{6}).

\section{Chromoelectric dipole moment of the top quark\label{2}}
The chromoelectric dipole moment $\tilde{d}^C$ is defined in the effective dimension $5$ operator
\beq
{\cal{L}}_I=-\frac{i}{2} \tilde{d}^C \bar{q} \sigma_{\mu \nu}\gamma_5 T^a q G^{\mu \nu a},
\eeq 
where $T^a$ are the $SU(3)$ generators and $G^{\mu\nu a}$ is the gluon field strength.
The contribution of this operator to the EDM of quarks can be computed using dimensional
analysis\cite{Manohar:1983md}. 
This technique can be expressed using the ``reduced" coupling
constant rule. Thus  the contribution of chromoelectric dipole moment operator to the EDM of the
quarks is given as follows
\beq
d^C=\frac{e}{4\pi} \tilde{d}^C,
\eeq
The alternative technique to estimate contributions of the chromoelectric operator is to
use the QCD sum rules\cite{Khriplovich:1996gk}. 
 We  note that the analysis of the top EDM is more complicated relative to EDM of the 
light quarks and of the light leptons (see e.g.,\cite{Ibrahim:1997nc,Ibrahim:1997gj}) because we cannot
ignore the mass of the external fermion (i.e., of the top quark in this case) compared to the masses that run
inside the loops. 
So the form factors that enter the analysis of the top EDM  are more complicated relative to 
the form factors that enter the EDM of the light quarks, since for the case of the top the
loop integrals are functions of more than just one mass ratio.

 \begin{figure*}[h]
 \begin{center}
      \includegraphics[width=12cm,height=4cm]{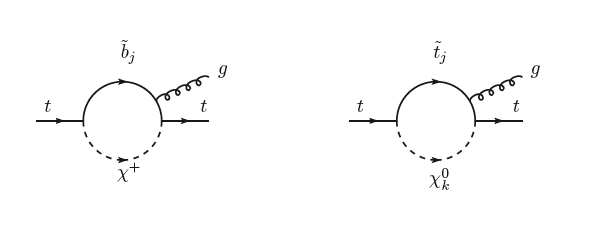} 
\caption{
Left: One loop contribution to the chromoelectric dipole moment of the top 
quark from the exchange of the chargino and from the exchange of sbottoms and 
mirror sbottoms.  Right: Same as the left diagram except that one has chromoelectric
dipole moment arising from the exchange of the neutralinos and from the exchange 
of stops and mirror stops. 
}
\label{fig1}
\end{center}
\end{figure*}

\section{Top CEDM from exchange of vector like multiplets\label{3}}
Using the formalism of \cite{Ibrahim:2010hv}, one can compute the contributions to the chromoelectric dipole
moment of the top quark. There are several contribution to it arising from the exchange of the
charginos, of the neutralinos, of the gluinos and of the W and Z boson. CP violation in these 
diagrams enters via the mass matrices involving the third  generation and their mirrors and similarly
via the mass matrices involving their superpartners and via the interaction vertices.  
A full description of the CP phases and the dependence of CEDM on them is given in Sec.(\ref{4}). 
We discuss now the various contributions to the CEDM of the top.

\subsection{Chargino exchange contribution}
The chargino exchange contribution to the chromoelectric dipole moment of the top quark arises through 
the left loop diagram of Fig.(\ref{fig1}). 
The relevant part of Lagrangian that 
 generates this contribution is given by 
\beqn
-{\cal{L}}_{t-\tilde{b}-\chi^+}=
\sum_{k=1}^2\sum_{i=1}^2\sum_{j=1}^4  
\bar{t}_k[\Gamma_{Lkji} P_L+ 
 \Gamma_{Rkji} P_R] 
\tilde{\chi^+}_i \tilde{b}_j +H.c.
\eeqn
where
\beqn
\Gamma_{Lkji}=-g[V^*_{i2}\kappa_t D^{t*}_{R1k}\tilde{D}^b_{1j}-D^{t*}_{R2k}V^*_{i1}\tilde{D}^b_{4j}+D^{t*}_{R2k}\kappa_B V^*_{i2}
\tilde{D}^b_{2j}],
\nonumber\\
\Gamma_{Rkji}=g[U_{i1} D^{t*}_{L1k}\tilde{D}^b_{1j}-D^{t*}_{L1k}\kappa_b U_{i2}\tilde{D}^b_{3j}-D^{t*}_{L2k}\kappa_T U_{i2}
\tilde{D}^b_{4j}],
\eeqn
where $\tilde{D}^b$ is the diagonalizing matrix of the  $4\times4$ sbottom mixed with scalar mirrors
 mass$^2$ matrix as defined 
in the appendix of \cite{Ibrahim:2010hv}.
These elements contain CP violating phases  can also contribute to the chromoelectric dipole moment of the top.
The couplings $\kappa_f$ are defined as 
\beqn
(\kappa_T, \kappa_{b})
=\frac{(m_T, m_{b})}{\sqrt{2} M_W \cos\beta},~
(\kappa_{B}, \kappa_{t})    =\frac{(m_B, m_{t})}{\sqrt{2} M_W \sin\beta}.
\eeqn
Here  $U$ and $V$ are the matrices  that  diagonalize the chargino mass matrix $M_C$ 
  so that 
\beq
U^* M_C V^{-1}= diag (m_{\tilde{\chi_1}}^+,m_{\tilde{\chi_2}}^+).
\eeq
Using the above interaction, we get from the left loop diagram  of Fig.(\ref{fig1})
 the contribution
\beqn
\tilde{d^C}(\chi^+)=\frac{g_s}{16\pi^2}\sum_{i=1}^2\sum_{j=1}^4 \frac{m_{\chi^+_i}}{m^2_{\tilde{b_j}}} Im(\Gamma_{L1ji}\Gamma^*_{R1ji})I_3(\frac{m^2_{\chi^+_i}}{m^2_{\tilde{b_j}}},\frac{m^2_{t_1}}{m^2_{\tilde{b_j}}}),
\eeqn
where $I_3(r_1,r_2)$ is given by
\beqn
I_3(r_1,r_2)=\int_0^1 dx \frac{x-x^2}{1+(r_1-r_2-1)x+r_2x^2},
\eeqn
We note that the limit of $I_3(r_1,r_2)$ for $r_2\sim 0$ is the well known form factors $B(r_1)$ 
in the case of light quarks \cite{Ibrahim:1997gj}. While our analysis is quite general  we will limit ourselves for simplicity 
to the case where
there is mixing between the third generation and the mirror part of the vector multiplet.
The inclusion of the non-mirror part is essentially trivial as it corresponds to an extension of the CKM
matrix from a $3\times 3$ to a $4\times 4$ matrix in the standard  model sector and
similar straightforward extensions in the supersymmetric sector. 
In the rest of the analysis we will focus just
on the mixings with the mirrors which is rather non-trivial.\\

\subsection{Neutralino exchange contribution}
The neutralino exchange contribution to the chromoelectric dipole moment of the top quark through 
the right loop diagram  of Fig.(\ref{fig1}). 
The relevant part of 
Lagrangian that generates this contribution is given by 
\beqn
-{\cal{L}}_{t-\tilde{t}-\chi^0}=
\sum_{k=1}^4\sum_{i=1}^4\sum_{j=1}^2  
\bar{t}_j[C_{Ljki} P_L+ 
C_{Rjki} P_R] 
\tilde{\chi^0}_i \tilde{t}_k +H.c.,
\eeqn
where
\beqn
C_{Ljki}=\sqrt{2}[\alpha_{ti}D^{t *}_{R1j} \tilde{D}^t_{1k}-\gamma_{ti}D^{t *}_{R1j} \tilde{D}^t_{3k}
+\beta_{Ti}D^{t *}_{R2j} \tilde{D}^t_{4k}-\delta_{Ti}D^{t *}_{R2j} \tilde{D}^t_{2k}],
\nonumber\\
C_{Rjki}=\sqrt{2}[\beta_{ti}D^{t *}_{L1j} \tilde{D}^t_{1k}-\delta_{ti}D^{t *}_{L1j} \tilde{D}^t_{3k}
+\alpha_{Ti}D^{t *}_{L2j} \tilde{D}^t_{4k}-\gamma_{Ti}D^{t *}_{L2j} \tilde{D}^t_{2k}].
\eeqn
The matrix  $\tilde{D}^t$ is the diagonalizing matrix of the $4\times 4$ stop mixed with scalar mirrors mass$^2$ matrix as defined in 
the appendix of \cite{Ibrahim:2010hv}.
The couplings that enter the above equations are given by
\beqn\label{alphabk}
\alpha_{t j} =\frac{g m_{t} X_{4j}}{2m_W\sin\beta},~~
\beta_{t j}=\frac{2}{3}eX_{1j}^{'*} +\frac{g}{\cos\theta_W} X_{2j}^{'*}
(\frac{1}{2}-\frac{2}{3}\sin^2\theta_W),\nonumber\\
\gamma_{tj}=\frac{2}{3}e X_{1j}^{'}-\frac{2}{3}\frac{g\sin^2\theta_W}{\cos\theta_W}
X_{2j}^{'},
~~ \delta_{t j}=-\frac{g m_{t} X_{4j}^*}{2m_W \sin\beta}.
\eeqn
Here 
\beqn
\alpha_{T j} =\frac{g m_{T} X^*_{3j}}{2m_W\cos\beta},~~
\beta_{T j}=-\frac{2}{3}eX_{1j}^{'} +\frac{g}{\cos\theta_W} X_{2j}^{'}
(-\frac{1}{2}+\frac{2}{3}\sin^2\theta_W),\nonumber\\
\gamma_{Tj}=-\frac{2}{3}e X_{1j}^{'*}+\frac{2}{3}\frac{g\sin^2\theta_W}{\cos\theta_W}
X_{2j}^{'*},
~~ \delta_{T j}=-\frac{g m_{T} X_{3j}}{2m_W \cos\beta},
\eeqn
where
\beqn
X'_{1j}= (X_{1j}\cos\theta_W + X_{2j} \sin\theta_W), 
~X'_{2j}=  (-X_{1j}\sin\theta_W + X_{2j} \cos\theta_W), 
\eeqn
and where the matrix $X$ diagonlizes the neutralino mass matrix so that
\beq
X^T M_{\tilde{\chi}^0} X=diag(m_{{\chi^0}_1}, m_{{\chi^0}_2}, m_{{\chi^0}_3}, m_{{\chi^0}_4}).
\eeq
Using the above interaction, we get from the right loop diagram  Fig.(\ref{fig1})
 the neutralino contributions to the top  chromoelectric dipole moment to be
\beqn
\tilde{d^C}(\chi^0)=\frac{g_s}{16\pi^2}\sum_{i=1}^4\sum_{k=1}^4 \frac{m_{\chi^0_i}}{m^2_{\tilde{t_k}}} Im(C_{L1ki} C^*_{R1ki})I_3(\frac{m^2_{\chi^0_i}}{m^2_{\tilde{t_k}}},\frac{m^2_{t_1}}{m^2_{\tilde{t_k}}}).
\eeqn
\subsection{Gluino exchange contribution}

\begin{figure*}[htb]
 \begin{center}
   \includegraphics[width=12cm,height=4cm]{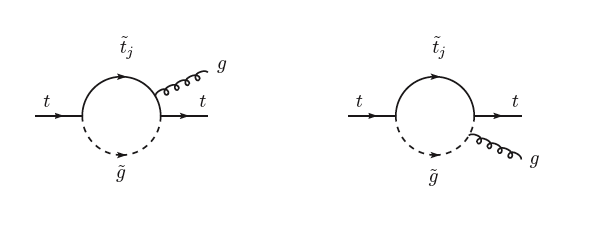} 
\caption{
Left: One loop contribution to the chromoelectric dipole moment of the top 
quark from  gluino exchange and from the exchange 
of stops and mirror stops. Here the external gluon line connects to  the stops and mirror stops
in the loop.  
Right: Same as the left diagram except that the external gluon line connects to  gluinos in the
loop, i.e., one has a gluino-gluino-gluon vertex in this case.
}
\label{fig2}
\end{center}
\end{figure*}

The gluino contribution to the chromoelectric dipole moment of the top  comes from the two loop diagrams of 
Fig.(\ref{fig2}). 
 The relevant part of Lagrangian that  generates this contribution is given by  
\beqn
-{\cal{L}}_{t\tilde{t} \tilde{g}}=\sqrt{2} g_s
\sum_{a=1}^8\sum_{j,k=1}^3\sum_{n=1}^2\sum_{m=1}^4 T^a_{jk} \bar t^j_{n}  
[K_{L_{nm}} P_L+K_{R_{nm}}P_R] \tilde{g}_a \tilde{t}^k_m +H.c. 
\eeqn
where
\beqn
K_{L_{nm}}=e^{-i\xi_3 /2}[D^{t*}_{R_{2n}} \tilde{D}^{t}_{4m}-D^{t*}_{R_{1n}}\tilde{D}^{t}_{3m}],\nonumber\\
K_{R_{nm}}=e^{i\xi_3 /2}[D^{t*}_{L_{1n}} \tilde{D}^{t}_{1m}-D^{t*}_{L_{2n}}\tilde{D}^{t}_{2m}],
\eeqn
where $\xi_3$ is the phase of the 
 gluino mass.
 
The above Lagrangian gives a contribution  
\beqn
\tilde{d^C}(\tilde{g})=\frac{g_s \alpha_s}{12\pi}\sum_{j=1}^4 \frac{m_{\tilde{g}}}{m^2_{\tilde{t_j}}} 
Im(K_{L_{1j}} K^*_{R_{1j}})I_5(\frac{m^2_{\tilde{g}}}{m^2_{\tilde{t_j}}},\frac{m^2_{t_1}}{m^2_{\tilde{t_j}}}),
\eeqn
where $I_5(r_1,r_2)$ is given by
\beqn
I_5(r_1,r_2)=\int_0^1 dx \frac{x+8x^2}{1+(r_1-r_2-1)x+r_2x^2}.
\eeqn
We note that the limit of $I_5(r_1,r_2)$ for $r_2\sim 0$ is the well known form factors $3C(r_1)$ 
in the case of light quarks \cite{Ibrahim:1997gj}.

\subsection{W and Z exchange contributions}

\begin{figure*}[t!]
 \begin{center}
   \includegraphics[width=12cm,height=4cm]{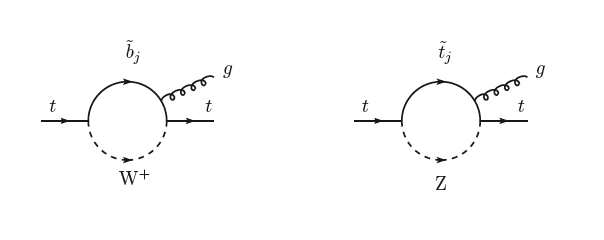} 
\caption{
Left: One loop contribution to the chromoelectric dipole moment of the top 
quark from $W^+$  exchange and from the exchange of the bottom quark and from the 
 mirror bottom.  Right: Same as the left diagram except that one has chromoelectric
dipole moment arising from Z exchange  and from the exchange 
of the top  and from the mirror top. 
}
\label{fig3}
\end{center}
\end{figure*}

The W boson exchange contribution to the chromoelectric dipole moment of the top quark  arises
through the left loop diagram of Fig.(\ref{fig3}).
 The relevant part of Lagrangian that 
 generates this contribution is given by 
\beqn
{\cal{L}}_{CC}=-\frac{g}{\sqrt 2} W^{+}_{\mu}
\sum_{i}\sum_{j} \bar t_{j} \gamma^{\mu} 
[D^{t *}_{L1j} D^{b }_{L1i} P_L+ 
 D^{t *}_{R2j} D^{b}_{R2i} P_R] 
 b_{i} +H.c. 
\label{LR}
\eeqn
where $i,j$ run over the set of quarks and mirror quarks including those from the third 
generation and from the vector multiplet,  
 $t_1$ is the physical top quark, and
$D^{t, b}_{L,R}$ are the diagonalizing  matrices defined in the appendix of \cite{Ibrahim:2010hv}.
 These matrices contain phases, and these phases
generate the chromoelectric dipole moment of the top quark.
Using the above interaction, we get from the left loop diagram of Fig.(\ref{fig3}), the contribution
\beqn
\tilde{d^C}(W^+)=\frac{g_s}{16\pi^2M^2_W}\sum_{i=1}^2m_{b_i} 
Im(\Gamma^{tb}_i)I_1(\frac{m^2_{b_i}}{M^2_W},\frac{m^2_{t_1}}{M^2_W}).
\label{EDM1}
\eeqn 
Here $\Gamma^{tb}_i$ is given 
\beq
\Gamma^{tb}_i=\frac{g^2}{2}D^{t *}_{L11} D^{b }_{L1i} D^{t}_{R21} D^{b *}_{R2i},
\eeq
and $I_1(r_1,r_2)$ is given by
\beqn
I_1(r_1,r_2)=\int_0^1 dx \frac{(4+r_1-r_2)x-4x^2}{1+(r_1-r_2-1)x+r_2x^2}.
\eeqn
 Finally we consider the right  loop of Fig.(\ref{fig3}) which produces the 
chromoelectric dipole moment of the top quark through the interaction with the Z 
boson.  The relevant part of Lagrangian that 
 generates this contribution is given by 
\beqn
{\cal{L}}_{NC}=-Z_{\mu}
\sum_{i=1}^2\sum_{j=1}^2\bar t_{j} \gamma^{\mu} 
[S_{Lji} P_L+ 
 S_{Rji} P_R] 
 t_{i}, 
\eeqn
where
\beqn
S_{Lji}=-\frac{g}{6\cos\theta_W}[-3D^{t*}_{L1j}D^t_{L1i}+4\sin^2\theta_W(D^{t*}_{L1j}D^t_{L1i}+D^{t*}_{L2j}D^t_{L2i})],
\nonumber\\
S_{Rji}=-\frac{g}{6\cos\theta_W}[-3D^{t*}_{R2j}D^t_{R2i}+4\sin^2\theta_W(D^{t*}_{R1j}D^t_{R1i}+D^{t*}_{R2j}D^t_{R2i})].
\eeqn
Using the above interaction, we get from the right loop of  Fig.(\ref{fig3}), the contribution
\beqn
\tilde{d^C}(Z)=\frac{g_s}{16\pi^2M^2_Z}\sum_{i=1}^2m_{t_i} Im(S_{L1i}S^*_{R1i})I_1(\frac{m^2_{t_i}}{M^2_Z},\frac{m^2_{t_1}}{M^2_Z}).
\label{EDM2}
\eeqn 
The total chromoelectric dipole moment of the top in the model is then given by the sum of the contributions
computed in this section so that 
\beqn
\tilde  d^C= \tilde d^C(\chi^+) +\tilde d^C(\chi^0) +\tilde d^C(\tilde g) + \tilde d^C(W^+) + \tilde d^C(Z).  
\eeqn

\section{Parameter space of the model and CP phases\label{4}}
The mass matrices for quarks and mirrors including their mixings are diagonalized using bi-unitary 
transformations $D^b_L$ and $D^b_R$ for the  bottom quarks and mirrors 
 and $D^t_L$ and $D^t_R$ for the diagonalization of the top quarks and mirrors.  We  parametrize $D^t_L$ and $D^t_R$ as follows 
\beqn
D^{t}_L=
 {\left(
\begin{array}{cc}
\cos\theta_L & -\sin\theta_L e^{-i\chi_L} \cr
             \sin\theta_L  e^{i\chi_L}& \cos\theta_L
\end{array}\right)},
~D^{t}_R=
 {\left(
\begin{array}{cc}
\cos\theta_R & -\sin\theta_R e^{-i\chi_R} \cr
             \sin\theta_R  e^{i\chi_R}& \cos\theta_R
\end{array}\right)}.
\eeqn
Thus the mixing between $t$ and $T$ is parameterized  by the angles $\theta_L$, $\theta_R$, $\chi_L$ and $\chi_R$
where the angles $\theta_L$, $\theta_R$ are given by
\beqn
\tan 2\theta_L=\frac{2|m_th^*_5-m_T h_3|}{m^2_t+|h_3|^2-m^2_T-|h_5|^2},
~\tan 2\theta_R=\frac{2|-m_th_3+m_T h^*_5|}{m^2_t+|h_5|^2-m^2_T-|h_3|^2},\nonumber\\
\eeqn
and $\chi_L$ and $\chi_R$ are the CP violating phases defined by
\beqn
\chi_R=arg (-m_{t} h_3+m_T h^*_5),
~\chi_L=arg (m_{t}h^*_5-m_T h_3). 
\eeqn
Similarly $D_L^b$ and $D_R^b$ are given by 
\beqn
D^{b}_L=
 {\left(
\begin{array}{cc}
\cos\theta_L & -\sin\phi_L e^{-i\xi_L} \cr
             \sin\phi_L  e^{i\xi_L}& \cos\phi_L
\end{array}\right)},
~D^{b}_R=
 {\left(
\begin{array}{cc}
\cos\phi_R & -\sin\phi_R e^{-i\xi_R} \cr
             \sin\phi_R  e^{i\xi_R}& \cos\phi_R
\end{array}\right)},
\eeqn
where the mixing between $b$ and $B$ is parametrized by the angle $\phi_L$, $\phi_R$, $\xi_L$ and $\xi_R$.
Here the angles  $\phi_L$ and $\phi_R$ are given by
\beqn
\tan 2\phi_L=\frac{2|m_bh^*_4+m_B h_3|}{m^2_b+|h_3|^2-m^2_B-|h_4|^2},
~\tan 2\phi_R=\frac{2|m_bh_3+m_B h^*_4|}{m^2_b+|h_4|^2-m^2_B-|h_3|^2}
\eeqn
and the  phases $\xi_{L,R}$ arise from the couplings $h_4$ and $h_3$ through 
the relations
\beqn
~\xi_R=arg (m_{b}h_3+m_B h^*_4),
~\xi_L=arg (m_{b}h^*_4+m_B h_3).
\eeqn
For the case of top and bottom masses arising from hermitian matrices, i.e., when
 $h_5=-h^*_3$ and $h_4=h^*_3$ we have
$\theta_L=\theta_R$, $\phi_L=\phi_R$, $\chi_L=\chi_R=\chi$ and $\xi_L=\xi_R=\xi$.  Further,   here we have
the relation $\xi=\chi+\pi$ and thus the W-exchange and the Z-exchange terms in the EDM for the top  
vanish.  However, more generally  the top and the bottom mass matrices are not hermitian and  they generate 
non-vanishing  contributions to the EDMs.
Thus the input parameters for this sector of the parameter space are
$m_{t1},m_T, h_3, h_5, m_{b1}, m_B, h_4$
with $h_3$, $h_4$ and $h_5$ being  complex masses with the corresponding 
CP violating phases $\chi_3$, $\chi_4$ and 
$\chi_5$.
For the sbottom and stop mass$^2$ matrices we need the extra input parameters of the susy breaking sector,
$
\tilde{M}_q, \tilde{M}_B,\tilde{M}_{b},\tilde{M}_{Q},\tilde{M}_{t},\tilde{M}_T,
A_{b}, A_T, A_{t},A_{B}, \mu, \tan\beta.
$
The chargino, neutralino and gluino sectors need the extra parameters
$\tilde{m}_1, \tilde{m}_2$ and $m_{\tilde{g}}$.
We will assume that the only parameters that have phases in the above set are 
$A_T$, $A_B$, $A_{t}$ and $A_{b}$ with the corresponding phases given by 
 $\alpha_T$, $\alpha_B$, $\alpha_{t}$ and $\alpha_{b}$.\\
 
 \begin{figure*}[t!]
 \begin{center}
   \includegraphics[scale=0.28]{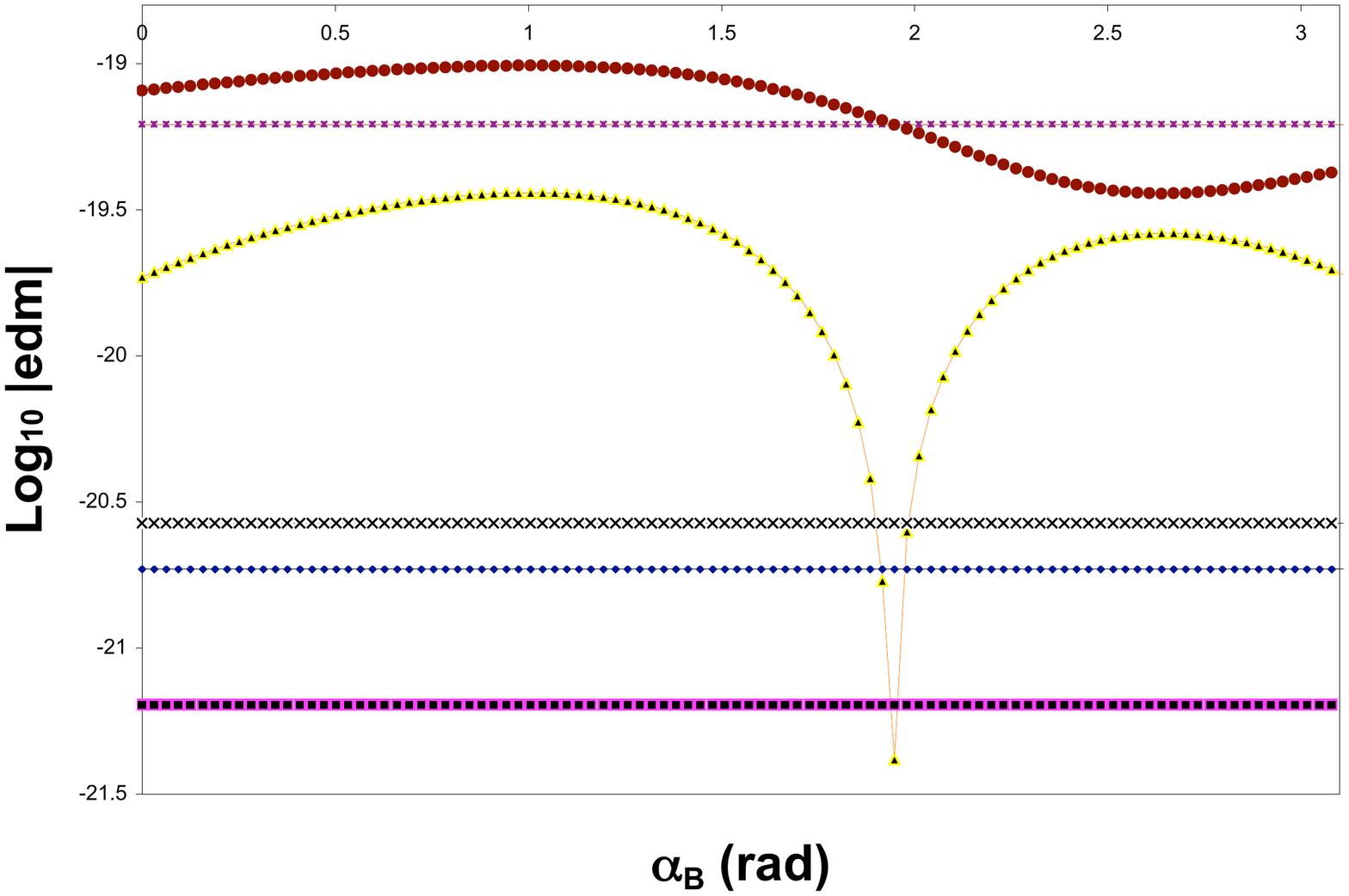} 
    \includegraphics[scale=0.28]{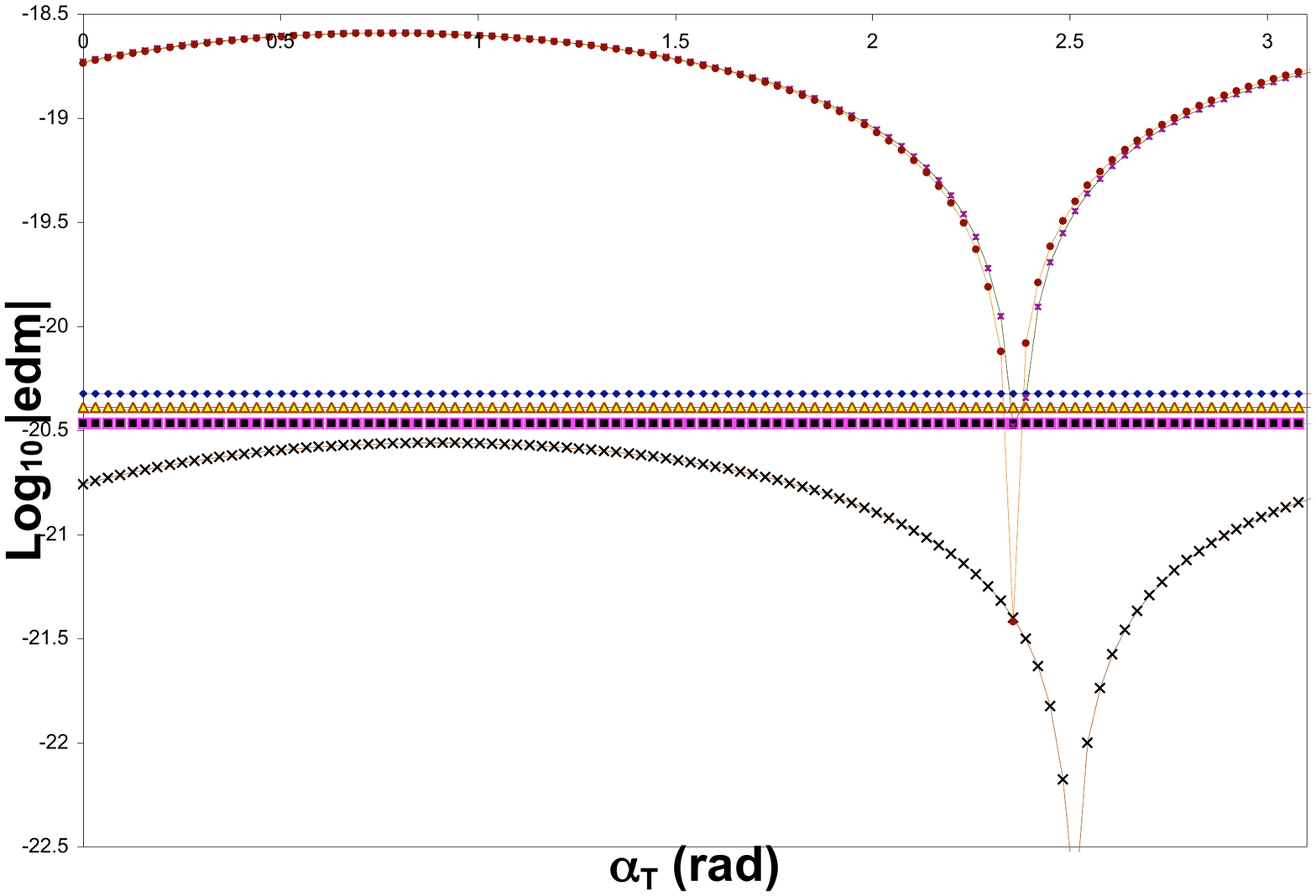} 
     \caption{\scriptsize
(Color online) 
Left:   An exhibition of the dependence of $d_{t}$ on $\alpha_B$ when $\tan\beta=5$, 
$m_T=250$, $|h_3|=$70, $|h_4|=$80, $m_B=$120,
$|h_5|=$90, $m_0=$220, $|A_0|=$200, $\tilde{m}_1=50$, $\tilde{m}_2=100$, 
$\mu=150$, $\tilde{m}_g=350$, $\chi_4=$0.3, $\chi_5=-$0.8, $\alpha_T=$0.4,   and $\chi_3=$0.4.
(The six curves correspond to the contributions from
 the  Z, W, neutralino, chargino, gluino and total CEDM. They are shown in ascending order at $\alpha_B=0$). 
Here and in subsequent figures all
masses are in GeV and  all angles are in rad.          
Right: An exhibition of the dependence of $d_{t}$ on $\alpha_T$ when $\tan\beta=25$, 
$m_T=200$, $|h_3|=$85, $|h_4|=$75, $m_B=$150,
$|h_5|=$85, $m_0=$200, $|A_0|=$200, $\tilde{m}_1=50$, $\tilde{m}_2=100$, 
$\mu=150$, $\tilde{m}_g=400$, $\chi_4=$0.5, $\chi_5=$0.7, $\chi_3=$0.8,   and $\alpha_B=$0.2.
(The six curves correspond to the contributions from
 the neutralino, Z, chargino, W, total CEDM and gluino. They are shown in ascending order at $\alpha_T=0$). 
}
\label{figa}
\end{center}
\end{figure*}

 \begin{figure*}[t!]
 \begin{center}
     \includegraphics[scale=0.28]{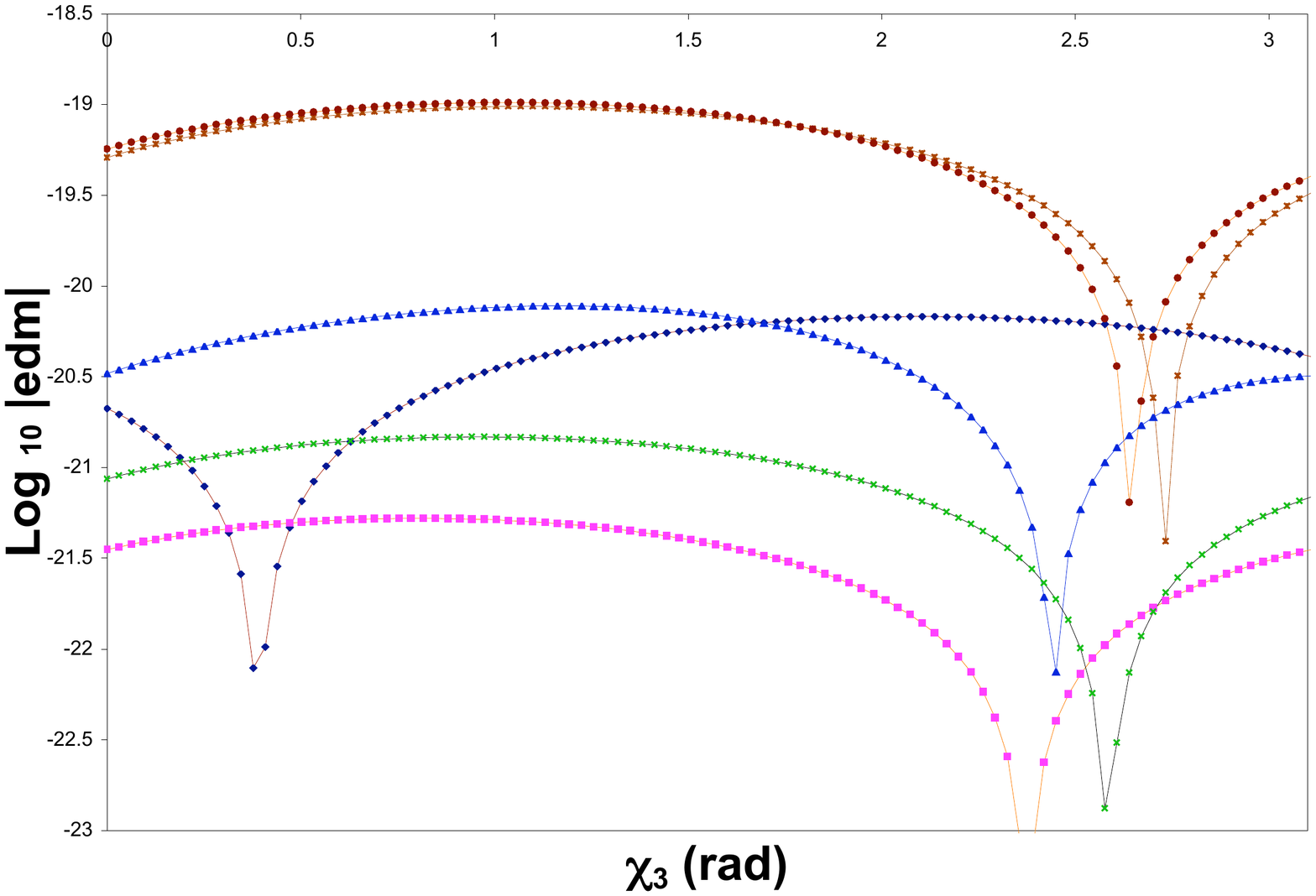} 
       \includegraphics[scale=0.28]{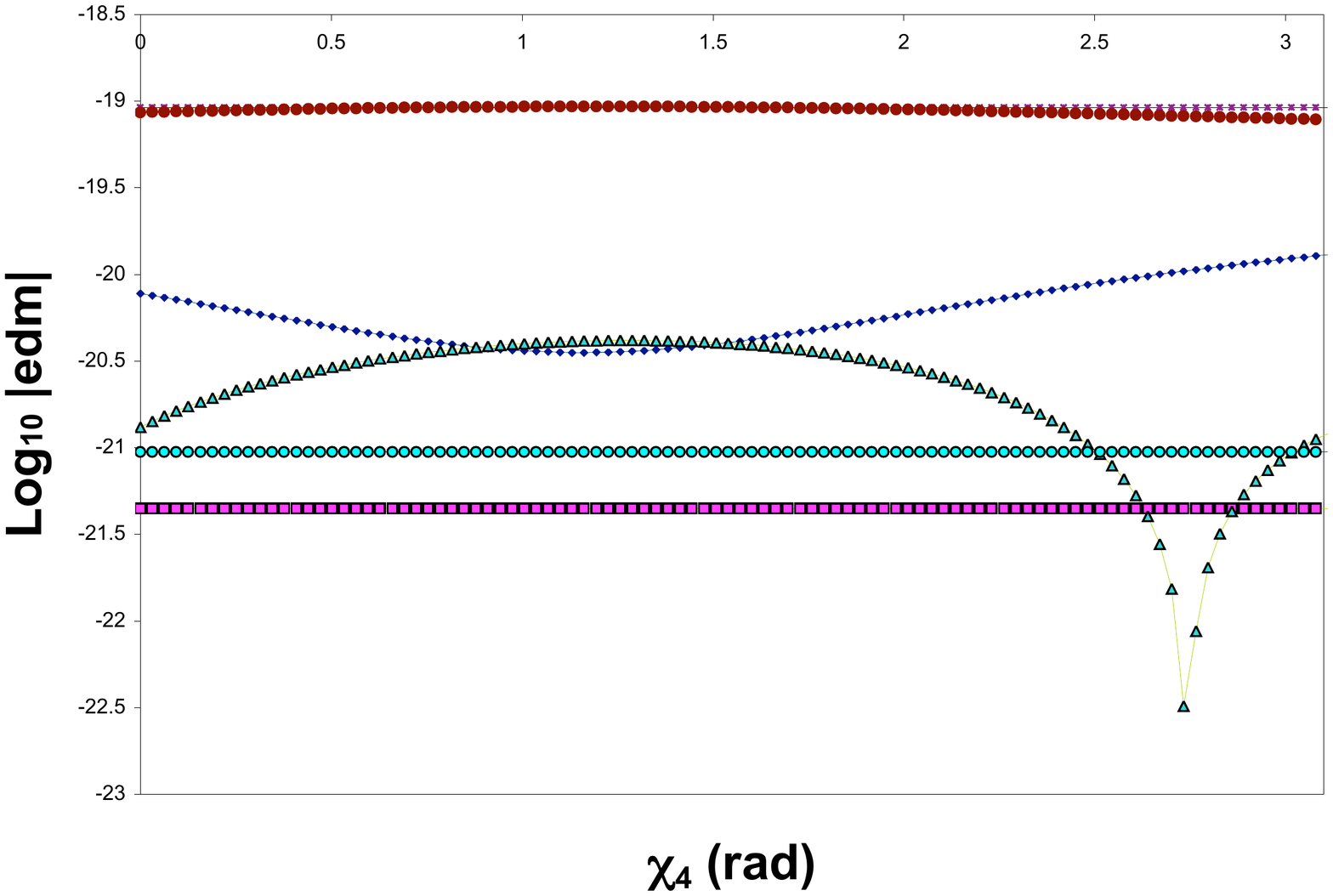} 
     \caption{\scriptsize
(Color online) 
Left:An exhibition of the dependence of $d_{t}$ on $\chi_3$ when $\tan\beta=10$, 
$m_T=150$, $|h_3|=$75, $|h_4|=$90, $m_B=$180,
$|h_5|=$80, $m_0=$300, $|A_0|=$300, $\tilde{m}_1=50$, $\tilde{m}_2=100$, 
$\mu=150$, $\tilde{m}_g=400$, $\chi_4=$0.7, $\chi_5=-$0.4, $\alpha_T=$0.2,   and $\alpha_B=$0.7.
(The six curves correspond to the contributions from
 the  Z, neutralino, W, chargino, gluino and total CEDM. They are shown in ascending order at $\chi_3=0$). 
Right:An exhibition of the dependence of $d_{t}$ on $\chi_4$ when $\tan\beta=15$, 
$m_T=350$, $|h_3|=$80, $|h_4|=$70, $m_B=$200,
$|h_5|=$100, $m_0=$400, $|A_0|=$400, $\tilde{m}_1=50$, $\tilde{m}_2=100$, 
$\mu=150$, $\tilde{m}_g=300$, $\chi_3=$0.6, $\chi_5=$0.8, $\alpha_T=$0.7,   and $\alpha_B=$0.2.
(The six curves correspond to the contributions from
 the Z, neutralino, chargino, W, total CEDM and gluino. They are shown in ascending order at $\chi_4=0$). 
}
\label{figb}
\end{center}
\end{figure*}

 \begin{figure*}[t!]
 \begin{center}
                 \includegraphics[scale=0.28]{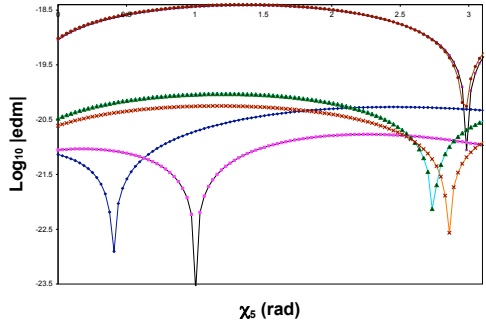} 
\caption{\scriptsize
(Color online) 
An exhibition of the dependence of $d_{t}$ on $\chi_5$ when $\tan\beta=20$, 
$m_T=300$, $|h_3|=$90, $|h_4|=$85, $m_B=$250,
$|h_5|=$95, $m_0=$100, $|A_0|=$200, $\tilde{m}_1=50$, $\tilde{m}_2=100$, 
$\mu=150$, $\tilde{m}_g=300$, $\chi_4=-$0.6, $\chi_3=$0.4, $\alpha_T=$0.7,   and $\alpha_B=$0.4.
(The six curves correspond to the contributions from
 the W, Z, neutralino, chargino, gluino and total CEDM. They are shown in ascending order at $\chi_5=0$).~~~~~~~~~~~~~~~~~~ 
}
\label{figc}
\end{center}
\end{figure*} 
 \section{Numerical estimate of the CEDM of the top\label{5}}
 To simplify the analysis further we set some of the  phases to zero, i.e., specifically we set 
$\alpha_{t}=\alpha_{b}=0$.
 With this in mind
the only contributions to the chromoelectric dipole moment CEDM of the top quark arises from mixing terms between the scalars and the
 mirror scalars, between the fermions - and the mirror fermions and finally among the mirror scalars themselves.
Thus in the absence of the mirror part of the lagrangian, the top CEDM  vanishes and so we can isolate
the  role of the CP violating phases in this sector and see the size of its contribution.
The $4\times 4$ mass$^2$ matrices of stops and sbottoms are diagonlized numerically.
Thus the CP violating phases that would play a role in this analysis are
\beq
\chi_3, \chi_4, \chi_5, \alpha_T, \alpha_B.
\eeq
To reduce the number of input parameters we assume
$\tilde{M}_a =m_0, ~a=q, B, b, Q, T, t$ and $|A_i|=|A_0|$, $i=T, B, t, b$.
In  the  left panel of  Fig(\ref{figa}), we  give a numerical analysis of the top EDM  and discuss
its  variation with the phase $\alpha_B$. 
We note that the only component that varies with this phase is the chargino component. This is expected since
$\alpha_B$ 
 enters the scalar bottom mass$^2$ matrix and  the chargino
contribution to the EDM is  controlled by   $\tilde{D}^b$ which depends on $\alpha_B$ while  the other 
contributions are independent of this phase.
Further, the chargino component exhibits a minimum where the different terms of it can have destructive 
cancellation.
In the  right panel of Fig(\ref{figa}), we study the variation of the different components of  $d_t$ on the  phase $\alpha_T$.
We observe that the  components that vary with this phase are the neutralino and the gluino contributions while
the W, Z and chargino contributions have no dependence on  this phase. The reason for the above is 
that  $\alpha_T$  enters
the  scalar top mass$^2$ matrix and the EDM arising from  W, Z and chargino exchanges are 
independent of $\tilde{D}^t$. 
However, 
 the neutralino and the gluino contributions are affected by it.
It is clear that we see here  the cancellation mechanism\cite{Ibrahim:1997nc,Ibrahim:1997gj,incancel,lpr}. 
working since the components are close to each other
with different signs, so we have the possibility of a destructive cancellation.

In the  left panel of Fig(\ref{figb}), we show the behavior of the different components of the chromoelectric dipole
moment contributions to the top EDM as a function of the phase $\chi_3$. 
 We note that 
 $\chi_3$ enters $D^{t}$, $D^{b}$, $\tilde D^{t}$ and $\tilde D^{b}$ and as a consequences all diagrams
in Fig.(1), Fig. (2) and  Fig. (3) that contribute to the top EDM have a $\chi_3$ dependence. 
Further, the  
 various diagrams that contribute to the top
 EDM may add constructively or destructively as shown in the Z, W, neutralino and chargino contributions. 
In the case of destructive interference, we have large
cancellations again reminiscent of the cancellation mechanism for the EDM of the electron and for 
the neutron\cite{Ibrahim:1997nc,Ibrahim:1997gj,incancel,lpr}.
Of course the desirable larger contributions  for the top  EDM occur away
from the cancellation regions.  
In the  right panel of  Fig(\ref{figb}), we study the variation of the different components of $d_t$ as the magnitude 
of the phase $\chi_4$ varies. 
The sparticle masses and couplings in the bottom sector and thus  the top
  EDM arising  from the exchange of the W and the charginos  are
  sensitive to $\chi_4$ and thus only these two  contributions to the top EDM  
  have dependence on  this parameter.
In   Fig(\ref{figc}), we study the variation of the different components of  $d_t$ as the phase $\chi_5$ changes.
This phase enters  the top quark mass matrix and the scalar top mass$^2$ matrix and consequently 
 the matrices $D_{L,R}^t$ and $\tilde{D}^t$. Thus the contributions to the EDM of the top arising from the
 W, Z, neutralino, chargino and gluino exchanges  all have a dependence on $\chi_5$ as exhibited in
Fig(\ref{figc}). \\

A comparison between the contributions of the chromoelectric dipole moment operator of the top EDM and that of the
electric dipole moment operator\cite{Ibrahim:2010hv}, shows that they could be
 the same order of magnitude with like or unlike signs.
That would provide an extra element for constructive or destructive interference of EDM components. To exhibit 
this, we give in Table 1 
the values of EDM for the top quark coming from the electric dipole moment operator and the
chromoelectric dipole moment operator. The first entry of Table 1 shows a destructive interference
between the electric and the chromoelectric dipole moments while the last two entries show a 
constructive interference.  With constructive interference a value of the top EDM as large as
 $\sim  6\times 10^{-19}$ ecm in magnitude (see the middle entry) can be gotten. 
 It is very possible that a full search of the parameter space of phases can lead to a top
 EDM of size $O(10^{-18})$ ecm. 
 
\begin{center} \begin{tabular}{|c|c|c|c|c|c|c|}
\multicolumn{6}{c} {Table~1:  Electric and chromoelectric dipole operator
contributions. } \\
\hline
$\chi_3 (rad)$ & $\chi_4$  &$\chi_5$ & $\alpha_T$ &$\alpha_B$ & $d_t^E e.cm$ &$d_t^C e.cm$  \\
\hline
\hline
$.3$     &  $-.5$  &  $1.0$     &   $.8$ & $-.4$
&    $8.04\times 10^{-19}$ & $-9.8\times 10^{-19}$
\\
 \hline
$.8$     &  $.4$  &  $-1.5$     &   $-.6$ & $.3$
&    $-1.57\times 10^{-19}$ & $-4.6\times 10^{-19}$
\\
\hline
$-.3$     &  $1.5$  &  $.1$     &   $.5$ & $-1.2$
&    $-1.73\times 10^{-19}$ & $-9.4\times 10^{-20}$

\\
\hline
 \hline
\end{tabular}\\~\\
\label{tab:1}
\noindent
\end{center}
Table caption:  A sample illustration of the  electric and chromoelectric dipole operator 
contributions to the electric dipole moment of the  top quark.  The inputs are:
$m_T=350$, $|h_3|=$100, $|h_4|=$175, $m_B=$100,
$|h_5|=$190, $m_0=$200, $|A_0|=$200, $\tilde{m}_1=50$, $\tilde{m}_2=100$, 
$\mu=150$, $\tilde{m}_g=450$
and   $\tan\beta= $ 5 (top row), 30 (middle row), 40 (bottom row). 
All masses are in units of GeV and all angles are in radian.

Constraints on the top chromo EDM have been obtained using the combined CDF and 
D\O\ data and the CMS and ATLAS data on the total $t\bar t$ pair production 
cross section  in \cite{HIOKI:2011xx,Hioki:2010zu}.
Further, it is shown in \cite{Gupta:2009wu,Antipin:2008zx} that  with $10$fb$^{-1}$ of data at $\sqrt s=14$ TeV at
the LHC a $5\sigma$ statistical sensitivity to a top quark chromo electric dipole moment of about $5\times 10^{-18} g_s.cm$ can be  reached. 

  \section{Conclusion\label{6}}
  Currently  the physics at the TeV scale is largely unknown and  it is hoped that the LHC will provide us
  with an insight in this energy regime. It  is fully conceivable that this energy regime contains extra anomaly free
  vector  like quark multiplets which can mix with the third generation. In this work we analyze the
  effect of this mixing on the chromoelectric dipole moment of the top quark. In this case one finds that
  there are  contributions that arise from the exchange of the extra vector like multiplets  in the 
  loops. We specifically focus on the exchange of the mirrors since their exchange can produce more
  dramatic contributions. Several sets of diagrams were computed for this analysis. These include the
  chargino exchange, the neutralino exchange, the gluino exchange as well as exchange of the W 
  and the Z boson bosons. In the analysis new sources of CP violation enter. They arise from the 
  complex mixing parameters of the third generation with the mirrors and from the soft parameter 
  involving interactions of the third generation with the mirrors.  Numerical analysis shows that an
  EDM as large as $10^{-18}$ ecm can be obtained for the top quark from the electric and chromoelectric 
  dipole contributions.  
   An EDM of this size could be accessible in future experiments such as at the ILC.

\noindent
{\em Acknowledgments}:  
We thank German Valencia for bringing to our attention the works of \cite{Gupta:2009wu,Antipin:2008zx}.
This research is  supported in part by  NSF grant PHY-0757959 and PHY-0704067.


\begin{thebibliography}{999}

\bibitem{Ibrahim:2007fb}
  T.~Ibrahim and P.~Nath,
  Rev.\ Mod.\ Phys.\  {\bf 80}, 577 (2008);
  arXiv:hep-ph/0210251;
 J.~R.~Ellis, J.~S.~Lee and A.~Pilaftsis,
  JHEP {\bf 0810}, 049 (2008)
  [arXiv:0808.1819 [hep-ph]];
  M.~Pospelov and A.~Ritz,
  Annals Phys.\  {\bf 318}, 119 (2005)
  [arXiv:hep-ph/0504231].

\bibitem{Hoogeveen:1990cb}
  F.~Hoogeveen,
  Nucl.\ Phys.\  B {\bf 341} (1990) 322;
 M.~E.~Pospelov and I.~B.~Khriplovich,
  Sov.\ J.\ Nucl.\ Phys.\  {\bf 53} (1991) 638
  [Yad.\ Fiz.\  {\bf 53} (1991) 1030].

\bibitem{Soni:1992tn}
  A.~Soni and R.~M.~Xu,
  Phys.\ Rev.\ Lett.\  {\bf 69}, 33 (1992).

\bibitem{note}
The analysis of \cite{Hoogeveen:1990cb} was for the electron and the 
EDM of the top is obtained by scaling as noted in \cite{Soni:1992tn}.

\bibitem{Atwood:2000tu}
  D.~Atwood, S.~Bar-Shalom, G.~Eilam and A.~Soni,
  Phys.\ Rept.\  {\bf 347}, 1 (2001)
  [arXiv:hep-ph/0006032].

\bibitem{Ibrahim:2008gg}
  T.~Ibrahim and P.~Nath,
  Phys.\ Rev.\  D {\bf 78}, 075013 (2008);
  [arXiv:0806.3880 [hep-ph]];
  Nucl.\ Phys.\ Proc.\ Suppl.\  {\bf 200-202}, 161 (2010)
  [arXiv:0910.1303 [hep-ph]].

\bibitem{Ibrahim:2010va}
  T.~Ibrahim and P.~Nath,
  Phys.\ Rev.\  D {\bf 81}, 033007 (2010)
  [arXiv:1001.0231 [hep-ph]].

\bibitem{Ibrahim:2010hv}
  T.~Ibrahim and P.~Nath,
  Phys.\ Rev.\  D {\bf 82}, 055001 (2010)
  [arXiv:1007.0432 [hep-ph]].

\bibitem{Georgi:1979md}
  H.~Georgi,
  Nucl.\ Phys.\  B {\bf 156}, 126 (1979);
  F.~Wilczek and A.~Zee,
  Phys.\ Rev.\  D {\bf 25}, 553 (1982);
J. Maalampi, J.T. Peltoniemi, and M. Roos, PLB 220, 441(1989);
  J.~Maalampi and M.~Roos,
  Phys.\ Rept.\  {\bf 186}, 53 (1990);
  K.~S.~Babu, I.~Gogoladze, P.~Nath and R.~M.~Syed,
  Phys.\ Rev.\  D {\bf 74}, 075004 (2006):
  Phys.\ Rev.\  D {\bf 74}, 075004 (2006); 
  P.~Nath and R.~M.~Syed,
  Phys.\ Rev.\  D {\bf 81}, 037701 (2010).

 \bibitem{Senjanovic:1984rw}
G.~Senjanovic, F.~Wilczek and A.~Zee,
  Phys.\ Lett.\  B {\bf 141}, 389 (1984);

\bibitem{Jezabek:1994zv}
  M.~Jezabek and J.~H.~Kuhn,
  Phys.\ Lett.\  B {\bf 329}, 317 (1994);
  C.~A.~Nelson, B.~T.~Kress, M.~Lopes and T.~P.~McCauley,
  Phys.\ Rev.\  D {\bf 56}, 5928 (1997);
  V.~M.~Abazov {\it et al.}  [D0 Collaboration],
  Phys.\ Rev.\ Lett.\  {\bf 100}, 062004 (2008).

\bibitem{Barger:2006fm}
  V.~Barger, J.~Jiang, P.~Langacker and T.~Li,
  Int.\ J.\ Mod.\ Phys.\  A {\bf 22}, 6203 (2007).

\bibitem{Lavoura:1992qd}
  L.~Lavoura and J.~P.~Silva,
  Phys.\ Rev.\  D {\bf 47}, 1117 (1993).

\bibitem{Maekawa:1995ha}
  N.~Maekawa,
  Phys.\ Rev.\  D {\bf 52}, 1684 (1995).

\bibitem{Morrissey:2003sc}
  D.~E.~Morrissey and C.~E.~M.~Wagner,
  Phys.\ Rev.\  D {\bf 69}, 053001 (2004)
  [arXiv:hep-ph/0308001].

\bibitem{Choudhury:2001hs}
  D.~Choudhury, T.~M.~P.~Tait and C.~E.~M.~Wagner,
  Phys.\ Rev.\  D {\bf 65}, 053002 (2002)
  [arXiv:hep-ph/0109097].

\bibitem{Liu:2009cc}
  C.~Liu,
  Phys.\ Rev.\  D {\bf 80}, 035004 (2009)
  [arXiv:0907.3011 [hep-ph]].

\bibitem{Babu:2008ge}
  K.~S.~Babu, I.~Gogoladze, M.~U.~Rehman and Q.~Shafi,
  Phys.\ Rev.\  D {\bf 78}, 055017 (2008).

\bibitem{Martin:2009bg}
  S.~P.~Martin,
  Phys.\ Rev.\  D {\bf 81}, 035004 (2010)
  [arXiv:0910.2732 [hep-ph]];
  Phys.\ Rev.\  D {\bf 82}, 055019 (2010)
  [arXiv:1006.4186 [hep-ph]];
  Phys.\ Rev.\  D {\bf 83}, 035019 (2011)
  [arXiv:1012.2072 [hep-ph]].

\bibitem{Graham:2009gy}
  P.~W.~Graham, A.~Ismail, S.~Rajendran and P.~Saraswat,
  arXiv:0910.3020 [hep-ph].

\bibitem{Frey:1997sg}
  R.~Frey {\it et al.},
{\it In the Proceedings of 1996 DPF / DPB Summer Study on New Directions for High-Energy Physics (Snowmass 96), Snowmass, Colorado, 25 Jun - 12
Jul 1996, pp STC119}
  [arXiv:hep-ph/9704243].

\bibitem{Atwood:1992vj}
  D.~Atwood, A.~Aeppli and A.~Soni,
  Phys.\ Rev.\ Lett.\  {\bf 69}, 2754 (1992).

\bibitem{Poulose:1997xk}
  P.~Poulose and S.~D.~Rindani,
  Phys.\ Rev.\  D {\bf 57}, 5444 (1998)
  [Erratum-ibid.\  D {\bf 61}, 119902 (2000)]
  [arXiv:hep-ph/9709225].

\bibitem{Choi:1995kp}
  S.~Y.~Choi and K.~Hagiwara,
  Phys.\ Lett.\  B {\bf 359}, 369 (1995)
  [arXiv:hep-ph/9506430].

 \bibitem{as2}
  A.~Soni and R.~M.~Xu,
  Phys.\ Rev.\ D.\  {\bf 45}, 2405 (1992).

      \bibitem{Bartl:1997wf}
        A.~Bartl, E.~Christova, T.~Gajdosik and W.~Majerotto,
        Nucl.\ Phys.\ Proc.\ Suppl.\  {\bf 66}, 75 (1998)
        [arXiv:hep-ph/9709219].

\bibitem{Hollik:1998vz}
  W.~Hollik, J.~I.~Illana, S.~Rigolin, C.~Schappacher and D.~Stockinger,
  Nucl.\ Phys.\  B {\bf 551}, 3 (1999)
  [Erratum-ibid.\  B {\bf 557}, 407 (1999)]
  [arXiv:hep-ph/9812298].

\bibitem{NovalesSanchez:2009zz}
  H.~Novales-Sanchez and J.~J.~Toscano,
  AIP Conf.\ Proc.\  {\bf 1116}, 443 (2009).

\bibitem{Huang:1994zg}
  C.~S.~Huang and T.~J.~Li,
  Z.\ Phys.\  C {\bf 68}, 319 (1995).

\bibitem{Manohar:1983md}
  A.~Manohar, H.~Georgi,
  Nucl.\ Phys.\  {\bf B234}, 189 (1984).

\bibitem{Khriplovich:1996gk}
  I.~B.~Khriplovich, K.~N.~Zyablyuk,
  Phys.\ Lett.\  {\bf B383}, 429-433 (1996).
  [hep-ph/9604211].


\bibitem{Ibrahim:1997nc}
  T.~Ibrahim and P.~Nath,
  Phys.\ Lett.\  B {\bf 418}, 98 (1998)
  [arXiv:hep-ph/9707409].

 \bibitem{Ibrahim:1997gj}
 T.~Ibrahim and P.~Nath,
  Phys.\ Rev.\  D {\bf 57}, 478 (1998);


  \bibitem{incancel}
   T. Ibrahim and P. Nath,  
   Phys. Rev. {\bf D58}, 111301(1998);
  Phys.\ Rev.\  D {\bf 58}, 111301 (1998);
   T.~Falk and K.~A.~Olive,
  Phys.\ Lett.\  B {\bf 439}, 71 (1998);
 M. Brhlik, G.J. Good, and G.L. Kane, Phys. Rev. {\bf D59}, 115004
 (1999); A. Bartl, T. Gajdosik, W. Porod, P. Stockinger, and
 H. Stremnitzer,  Phys. Rev. {\bf 60}, 073003(1999);
 S. Pokorski, J. Rosiek and C.A. Savoy, 
 Nucl.Phys. {\bf B570}, 81(2000);
 E.~Accomando, R.~Arnowitt and B.~Dutta,
Phys.\ Rev.\ D {\bf 61}, 115003 (2000);
  U. Chattopadhyay, T. Ibrahim, D.P. Roy, Phys.Rev.D64:013004,2001;
 C.~S.~Huang and W.~Liao,
Phys.\ Rev.\ D {\bf 61}, 116002 (2000);
ibid, Phys.\ Rev.\ D {\bf 62}, 016008 (2000);
 M. Brhlik, L. Everett, G. Kane and J. Lykken, Phys. Rev.
 Lett. {\bf 83}, 2124, 1999; Phys. Rev. {\bf D62}, 035005(2000);
T.~Ibrahim and P.~Nath,
  Phys.\ Rev.\  D {\bf 61}, 093004 (2000);
  T.~Ibrahim,
  Phys.\ Rev.\  D {\bf 64}, 035009 (2001);
 T. Falk, K.A. Olive, M. Prospelov, and R. Roiban, Nucl. Phys. 
 {\bf B560}, 3(1999); V.~D.~Barger, T.~Falk, T.~Han, J.~Jiang, T.~Li 
 and T.~Plehn,
Phys.\ Rev.\ D {\bf 64}, 056007 (2001);
T.~Ibrahim and P.~Nath,
  Phys.\ Rev.\  D {\bf 67}, 016005 (2003).

\bibitem{lpr}
  Y.~Li, S.~Profumo and M.~Ramsey-Musolf,
  JHEP {\bf 1008}, 062 (2010)
  [arXiv:1006.1440 [hep-ph]].


\bibitem{HIOKI:2011xx}
  Z.~Hioki and K.~Ohkuma,
  arXiv:1104.1221 [hep-ph].

\bibitem{Hioki:2010zu}
  Z.~Hioki and K.~Ohkuma,
  Eur.\ Phys.\ J.\  C {\bf 71}, 1535 (2011)
  [arXiv:1011.2655 [hep-ph]].

\bibitem{Gupta:2009wu}
  S.~K.~Gupta, A.~S.~Mete and G.~Valencia,
  Phys.\ Rev.\  D {\bf 80}, 034013 (2009)
  [arXiv:0905.1074 [hep-ph]] and Private Communication with German Valencia.
  
\bibitem{Antipin:2008zx}
  O.~Antipin and G.~Valencia,
  Phys.\ Rev.\  D {\bf 79}, 013013 (2009)
  [arXiv:0807.1295 [hep-ph]].
  


\end{thebibliography}
\end{document}